\documentclass[12pt]{article}
\usepackage{dcolumn}
\usepackage{graphicx}
\usepackage{amssymb}
\newcommand{\be}{\begin{eqnarray}}
\newcommand{\ee}{\end{eqnarray}}
\newcommand{\pl}[1]{Phys. Lett. {\bf #1}}

\newcommand{\prl}[1]{Phys. Rev. Lett. {\bf #1}}
\newcommand{\pr}[1]{Phys. Rev. {\bf #1}}

\newcommand{\np}[1]{Nucl. Phys. {\bf #1}}

\def\e{{\rm e}}

\def\inti{\int \!\!}

\begin{document}

\title {All orders breakup of heavy exotic nuclei in a semiclassical model}

\author{ A. Garc\'ia-Camacho $^{a}$,   A. Bonaccorso $^{a}$, D. M. Brink 
$^{b}$ \\ \small
     $^{(a)}$ Istituto Nazionale di Fisica Nucleare, Sezione di
Pisa,.\\   \small and Dipartimento di Fisica, Universit\`a di Pisa,\\\small via Largo Pontecorvo 3, 56127
Pisa, Italy.\\
\small $^{(b)}$ Department of Theoretical Physics, 1 Keble Road, Oxford OX1 3NP, U. K.\\
 }

\date{\today}
\maketitle
\begin{abstract}
We present a semiclassical and mostly analytical model of elastic neutron breakup reactions for exotic projectiles. Both nuclear and Coulomb induced reactions are considered and the potentials are treated to all orders in the interactions. Furthermore we introduce a technique which allows the use of the full Coulomb potential, thus including all multipoles besides the dipole. Results for deeply bound states as well as for halo initial states are presented and it is shown that experiments on heavy targets would be well suited to study exotic nuclei with tightly bound valence nucleons.
\end{abstract}

{\bf Pacs} {21.10.Jx, 24.10.-i, 25.60.Gc, 27.30.+t}

\maketitle

\section{Introduction}

        The usefulness of breakup reactions in addressing the problem of exotic nuclei structure is widely recognized since many years. In particular, the study of spectroscopy and single-particle strength of halo nuclei allows a direct comparison between theory and experiment, but in order to do so reliable reaction models are needed to obtain accurate cross-section calculations. Therefore a good understanding of the reaction mechanism is an issue of capital importance.

        This work regards one-neutron breakup from exotic nuclei, with both halo and tightly bound valence particles. We concentrate on reactions on heavy targets at intermediate beam energies. The projectile ground state is regarded as a single-particle state in the two-body system made of a valence neutron plus a heavy fragment.  Most of the experiments are inclusive and only the heavy fragment is detected, but sometimes exclusive experiments with  the valence neutron  detected in coincidence \cite{ann94}-\cite{gsi} with the core have been performed. Since we consider reactions involving heavy targets,  the process is Coulomb dominated for halo projectiles. However we will show that  for normally bound initial states Coulomb breakup becomes negligible and  breakup reactions can be used as a tool to obtain structure information provided nuclear breakup is  properly included. 
        
        There have been several approaches \cite{typ01a}-\cite{cha02} to this problem, some of which consist in  numerical solutions of the Schr\"odinger equation \cite{esben95}-\cite{capel04}. These approaches have reached a large degree of accuracy but the physics interpretation of the results partially relies on the comparison with results from  more analytical models. Furthermore due to the numerical complexity the scattering of two heavy nuclei has not been treated so far. Thus our focus is not so much on the numerical aspects but rather on the understanding of the main features of the reaction mechanism in a simple, yet accurate, picture. In this spirit, earlier works \cite{mar02,mar03} have shown the importance of treating the nuclear and Coulomb interactions in a consistent way and to all orders. 
        
        Within the semiclassical formalism, we now attempt to examine two aspects of the reaction process: the first one is the peripherality of the reaction, whereas the second concerns the accuracy of the dipole approximation for the Coulomb interaction. They are both important to extend breakup studies to heavy reaction partners. The concept of the reaction being peripheral is important as it allows the use of spectator-like models for the core target interaction leading to a semiclassical relative motion trajectory. It underlies a large part of existing approaches, and has been recently rediscussed by Capel et al.\cite{capel04}. The fact that the heavy fragment is detected, i.e. it survives the collision suggests that processes where it collides closely with the target would not contribute significantly to the measured cross-section. The main contribution is expected to come from interactions between the target and the tail of the valence neutron single-particle wave function. In halo nuclei this tail goes much beyond the range of the potential. One of the aims of this work is to compare the cross-sections obtained by using the whole single-particle wave function with that given just by its tail. This turns the formalism of \cite{mar02,mar03} less analytical and adds some numerical complications, but some light can be shed on the problem of which part of the projectile wave function the reaction is sensitive to. This is an interesting point being at the center of a scientific debate since recently  the reliability of spectroscopic factors obtained from peripheral reactions has been questioned \cite{mukha05}.
       For deeply bound states peripherality is best obtained on heavy targets. But as we will show in the following in these cases Coulomb and nuclear breakup can be  of the same order depending on the nucleon separation energy and they need to be treated consistently up to separation energies around 8 MeV. However for larger separation energies and angular momentum states $\ell>1$ Coulomb breakup becomes negligible. Therefore for reactions involving very neutron rich projectiles we recommend the use of heavy targets.
       
        Regarding our second aim, the formalism developed in \cite{mar02,mar03} included Coulomb potential only in its dipole form, although all orders in perturbation theory were considered. It was shown that full description of the time-evolution could be obtained analytically. On the other hand, other papers \cite{esben96} have stressed the importance of going beyond the dipole term, in particular when one deals with proton haloes. A series of works has also focused on the possible effects of dipole-quadrupole interference \cite{davids01}. The second goal of this paper is therefore to calculate the breakup due to the full Coulomb potential and to compare the results to those from the simple dipole approximation. This is done in preparation for a future study of proton breakup.
        
 The new aspects of the present approach with respect to our previous works \cite{mar02,mar03} are therefore: i) the use of the full numerical  Woods-Saxon wave function for the initial neutron state: ii) an all-multipoles-all-order  treatment of the Coulomb potential; iii) the application to reactions in which both projectile and target are heavy nuclei.

\section{Theoretical framework} \label{teo}

\begin{figure}
\center
\includegraphics[scale=0.5,width=7cm]{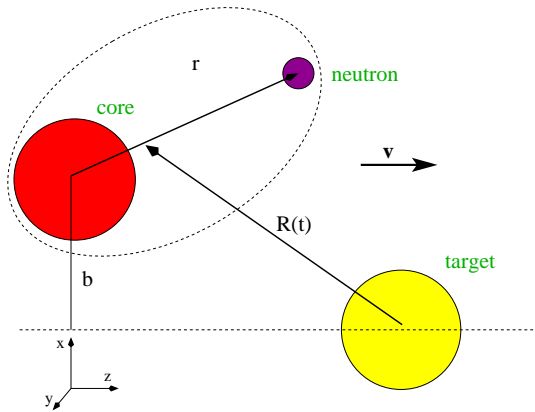}
\caption{Reference frame and coordinates used in this work.}
\label{pall}
\end{figure}

In previous papers \cite{mar02,mar03} some of us   introduced a full description of the treatment of the
scattering equation for a projectile which decays by single neutron breakup due to its
interaction with the target. 
  There it was shown that within the semiclassical approach for the
projectile-target relative motion, the amplitude for a
   transition from a  nucleon bound state $\psi_i$ in the projectile to a final
continuum state
$\psi_f$ is given by 
\begin{equation}
g_{fi}={1\over i\hbar}
\int_{-\infty}^{\infty}dt<\psi_{f} (\vec{r},t)|V(\vec{r},\vec{R}(t))|\psi_{i}(\vec{r},t)>,
\label{amp0}
\end{equation}
where $V$ is the neutron-target interaction, responsible for the neutron transition to the continuum.  $\vec{R}(t)$ is the core-target relative  coordinate, that is approximated by a straight line trajectory with constant velocity $v$, $R(t)=\hat{x} R_\perp + \hat{z} vt$, and $\vec{r}$ is the neutron-core coordinate. As usual in a semiclassical formalism, the concept of trajectory brings on the time dependence. Explicitly, the initial state wave function depends on time as
$
\psi_{i}(\vec{r},t)= e^{-i \varepsilon_0 t /\hbar}\phi_i(\vec{r}),
$
 where $\varepsilon_0$ is the binding energy of the valence neutron.
The final state wave function,   will be specified later.
For light targets the recoil effect due to the projectile-target Coulomb potential is
rather small  and the interaction responsible for the reaction is mainly the
neutron-target nuclear potential. In the case of heavy targets the dominant reaction
mechanism is Coulomb breakup.  The Coulomb force does not act directly on the neutron but
it affects it only indirectly by causing the recoil of the core.
 In Ref. \cite{mar02} it was shown that the combined
effect of the nuclear and Coulomb interactions to all orders  can be taken into account by
using the potential
${V}=V_{\rm  nt}+V_{\rm  eff}$ sum of the neutron-target optical potential and the effective Coulomb
dipole potential.  In Ref. \cite{mar02}, as in this paper we neglect the neutron-core final state interaction under the hypothesis that the observables calculated do not depend significantly on it \cite{noi}. If for the neutron final continuum  wave function we take a distorted
wave of the  eikonal-type
\begin{equation}
\psi_f (\vec{r}, t)=exp\left\{ i\vec{k} \cdot \vec{r} + i\chi_{\rm eik}(\vec{r}, t)- i\frac{\varepsilon_ft}{\hbar}\right\}
\end{equation}
 where $\vec{k}\equiv (k_x,k_y,k_z)$ is a real vector and the eikonal phase shift is
simply
\begin{equation}
\chi_{\rm  eik}(\vec{r},t)=\frac{1}{\hbar } \int_{t}^\infty V ( \vec{r},\vec{R}(t^\prime))
dt^\prime. \label{ekph}
\end{equation}
The coordinate system used in this work is presented in Fig.\ref {pall}, where
\begin{eqnarray}
b = R_\perp-r_\perp=R_\perp-\frac{\vec{R}_\perp \cdot \vec{r}}{R_\perp}
\end{eqnarray}

Then the amplitude in the projectile reference frame becomes
\begin{equation}
g_{fi}  ( \vec{k},\vec{b})  = \frac{1}{i\hbar}\inti
d^{3} \vec{r} \int dt \e^{-i{\vec k} \cdot  {\vec r}+i\omega t -i\chi_{\rm  eik}(\vec{r},t)}
V \left (  \vec{r},\vec{R}(t)\right )
  \phi_{i}\left(  \vec{r}\right)
\label{amp1}
\end{equation}
where $\phi_{i}$ is the time independent part of the neutron initial wave
function and   $i\equiv(l,m)$ stands for the angular momentum quantum numbers,
$\omega=\left(  {{\varepsilon  _f}}-\varepsilon_{0}\right)/\hbar
$ and $\varepsilon_0$ is the neutron initial bound state energy  while
${{\varepsilon  _f}}$ is the  final neutron-core continuum energy. This formalism is closely related to \cite{ald75}.
Eq.~(\ref{amp1}) is appropriate to calculate the coincidence cross section 
$A_p\to (A_p-1)+n$. Finally the
differential probability 
 to the neutron momentum can be
written as
\begin{equation}{d^3P_{bu}(d)\over d\vec{k}} ={1\over 8\pi^3}{1\over
2l_i+1}\Sigma_{m_i m_f} |g_{fi}(  \vec{k},\vec{b})|^2 .\label{anc}\end{equation} and we have averaged over the neutron
initial magnetic substates.

\subsection{Coulomb phase}
 The dipole approximation to the Coulomb potential is
\begin{eqnarray} 
V_{Cou}(\vec{r},\vec{R})= \frac{V_0}{|\vec{R}-\beta_1 \vec{r}|} -  \frac{V_0}{R} \simeq V_0 \beta_1 \frac{\vec{R} \cdot \vec{r}}{R^3} + O(r^2),
\end{eqnarray}
 where $V_0=Z_c Z_t e^2$ and $\beta_1=m_c/m_p$. The centre of mass of the projectile is assumed to follow a straight line, $R(t)=\hat{x} R_\perp + \hat{z} vt$ where  $v$ is the relative motion velocity at the distance of closest approach. 

As stated above, one of the aims of this paper is to derive analytical formulae for the breakup amplitude when the above approximation is not performed. As done in \cite{mar02}, this implies the calculation of the integral
\begin{eqnarray} \label{chichi}
 \chi_{eff}(\vec{b},\vec{r},k)=\frac{1}{\hbar}\int dt e^{i\omega t} V_{Cou}(\vec{r},t),
\end{eqnarray}
 with $\omega=(\epsilon_f-\epsilon_i)/\hbar$. Such integral is solvable in the dipole approximation. In order to obtain a similar result for the whole multipole expansion, a screening term is added and subtracted to the potential, thus it can be written as
\begin{eqnarray} \label{poti}
V(\vec{r},\vec{R})=V_{sh}(\vec{r},\vec{R})  + V_{lo}(\vec{r},\vec{R}),
\end{eqnarray}
 where
\begin{eqnarray}
V_{sh}(\vec{r},\vec{R})=V_0 \frac{e^{- \gamma |\vec{R}-\beta_1 \vec{r}|}}{|\vec{R}-\beta_1 \vec{r}|} - V_0 \frac{e^{- \gamma R}}{R},
\end{eqnarray}
\begin{eqnarray}
V_{lo}(\vec{r},\vec{R})=- V_0 \frac{1-e^{- \gamma |\vec{R}-\beta_1 \vec{r}|}}{|\vec{R}-\beta_1 \vec{r}|}+ V_0 \frac{1-e^{- \gamma R}}{R}.
\end{eqnarray}
 The term $V_{sh}$ contains the singularity at $R=0$ but decays quickly with the impact parameter. On the other hand, $V_{lo}$, well-behaved in the origin, accounts for the long-range character of the Coulomb potential. When inserted in eq. (\ref{chichi}), these two terms can be treated in different ways if the parameter $\gamma$ is big enough. In this case, as done in \cite{mar02} with the nuclear potential, $V_{sh}$ can be considered in the sudden approximation, yielding a phase
\begin{eqnarray}
\chi_{sudd}(\vec{b},\vec{r})=\frac{1}{\hbar}\int dt  V_{sh}(\vec{r},t),
\end{eqnarray} 
 whereas $V_{lo}$ needs to keep the whole time evolution description, but, being weak, it can be approximated to first order
\begin{eqnarray}
\chi_{pert}(\vec{b},\vec{r},k)=\frac{1}{\hbar}\int dt e^{i\omega t} V_{lo}(\vec{r},t).
\end{eqnarray} 
 Therefore the Coulomb phase becomes a sum of two terms
\begin{eqnarray}
 \chi_{eff}(\vec{b},\vec{r},k)= \chi_{sudd}(\vec{b},\vec{r})+ \chi_{pert}(\vec{b},\vec{r},k),
\end{eqnarray}
 both of them depending upon the screening parameter $\gamma$, as
\begin{eqnarray} \label{suph}
\chi_{sudd}(\vec{b},\vec{R})=\frac{V_0}{\hbar v}\left(K_0(\gamma b)-K_0(\gamma R_\perp)\right),
\end{eqnarray} 
\begin{eqnarray} \label{chff}
&&\chi_{pert}(\vec{b},\vec{r},k)= \frac{2V_0}{\hbar v} e^{i\beta_1 \omega z/v} \left( K_0(b\omega /v) - K_0(\sqrt{(b \omega/v)^2+(\gamma b)^2}) \right) \nonumber \\
&-& \frac{2V_0}{\hbar v}  \left(K_0(R_\perp \omega/v) - K_0(\sqrt{(R_\perp \omega/v)^2+(\gamma R_\perp)^2}) \right).
\end{eqnarray}

 In order for this approximation to be valid, the screening term $\gamma$ needs to be sufficiently large as to ensure that the range of $V_{sh}$ remains short enough, and that $V_{lo}$ does not become too large. This is next achieved by taking just $\gamma=\infty$, in which case
\begin{eqnarray}
\chi_{sudd}=0
\end{eqnarray}
\begin{eqnarray} \label{fiqui}
\chi_{pert}=\frac{2 V_0}{\hbar v}\left(e^{i\beta_1 \omega z /v}K_0(\omega b/v) -K_0(\omega R_\perp/v)\right)
\end{eqnarray}
 If eq. (\ref{fiqui}) is expanded up to first order in $\vec{r}$, we obtain
\begin{eqnarray} \label{fuqui}
\chi_{pert} \simeq \frac{2 V_0}{\hbar v}\left(K_0(\omega R_\perp/v)\frac{i\omega z}{v} \beta_1+K_1(\omega R_\perp/v)\frac{\vec{R}_\perp \cdot \vec{r}}{R_\perp}\frac{\omega}{v} \beta_1\right),
\end{eqnarray}
 consistent with the result of \cite{mar02}.

\subsection{Sudden limit and all-order treatment}

 Eq. (\ref{fiqui}) gives rise to a probability amplitude which contains only first order in the interaction potential $V_0$. Aiming for an all-order formalism, in \cite{mar03} it was shown that a possible way to achieve this is to use sudden approximation, subtract the first order term, which diverges for large impact parameter, and then to add a first order term calculated in time-dependent perturbation theory, i.e. eq. (\ref{fiqui}).

 The sudden limit ($\omega \to 0$) must be therefore taken in the above expression for $\chi_{eff}$, yielding 
 \begin{eqnarray} \label{suddi}
\chi_{eff}^{sudd}=\frac{2 V_0}{\hbar v} \log{\frac{b}{R_\perp}}.
\end{eqnarray}
 The Coulomb breakup probability amplitude is therefore
\begin{eqnarray}
g^{Cou}= \int d\vec{r} e^{i\vec{k} \cdot \vec{r}} \phi_i(\vec{r}) \left( e^{i\chi_{eff}^{sudd}} - 1 - \frac{2 V_0}{\hbar v} \log{\frac{b}{R_\perp}} + \chi_{pert} \right)
\end{eqnarray}
 Moreover, the nuclear part of the breakup amplitude is given by
\begin{eqnarray} \label{nuq}
g^{nuc}=\int d\vec{r} e^{i\vec{k}\cdot \vec{r}} \left( e^{i \chi_{nt}(b)}-1\right) \phi_{i}(\vec{r}).
\end{eqnarray}
 With these ingredients, the expression for the differential cross-section is just
\begin{eqnarray}
\frac{d \sigma}{d \epsilon}= \frac{1}{4\pi^2}\frac{m k}{\hbar^2}\int d b b |S_{ct}(b)|^2 |g^{nuc}+g^{Cou}|^2,
\end{eqnarray}
 where $S_{ct}(b)$ is the core-target S-matrix, depending upon its appropriate impact parameter $b$. Then we can apply this expressions of all-multipoles and whole wave function in two formalisms:

a) - \cite{mar02} Nuclear in sudden approximation and Coulomb to first order in $\chi_{eff}$.

b) - \cite{mar03} Nuclear and Coulomb in sudden approximation, first order sudden Coulomb subtracted and first order perturbation added.

\section{Applications} \label{app}

\subsection{Two-body interactions} \label{bou}

 In the calculations below, the ground state of $^{11}$Be is assumed to be a purely single particle $s$-state of the two-body $^{10}$Be-n system. Possible deformations of the $^{10}$Be core are not considered, thus its intrinsic spin is just zero in our approach. The ground state wave function is calculated in a Wood-Saxon potential with fixed geometry, radius 1.25$A_c^{1/3}$ fm and diffuseness 0.6 fm.

 Our formalism only considers the interaction between the core and the target inside the core 
 probability, given by the modulus square of the S-matrix S$_{ct}$. This S-matrix is calculated by making use of a potential V$_{ct}$ obtained by double-folding the matter densities of $^{10}$Be and $^{208}$Pb with an effective nucleon-nucleon interaction \cite{yomismo}. The densities where assumed to have gaussian shapes, with parameters chosen to fit the experimentally determined root mean square radii, which are 2.30 fm for $^{10}$Be \cite{oza01} and 5.6 fm for $^{208}$Pb \cite{kara02}.

 Finally, when nuclear breakup is addressed below, a potential interaction V$_{nt}$ between the valence particle and the target is required. The optical potential of \cite{jlm1,jlm} is used in our calculations.

\section{Spectrum of elastic breakup of $^{11}$Be on $^{208}$Pb}

\subsection{Validity of asymptotics}
\begin{figure}
\center
\includegraphics[scale=0.5,width=10cm]{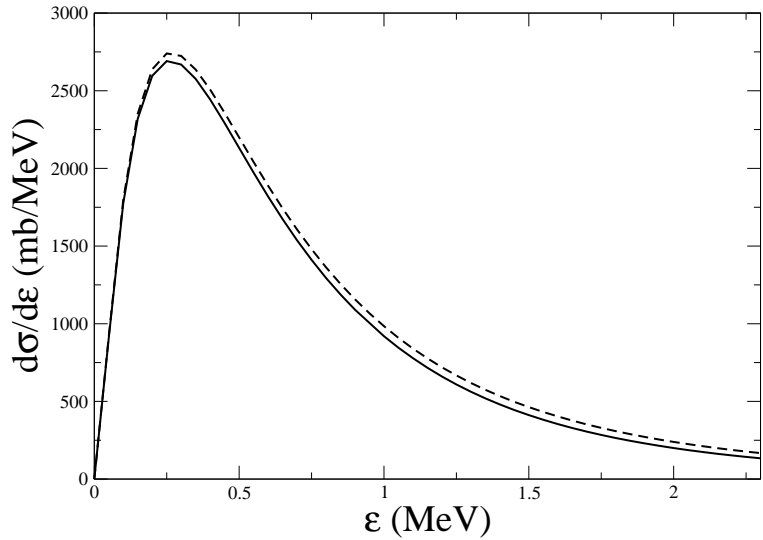}
\caption{Calculations of differential cross-section for Coulomb breakup of $^{11}$Be on $^{208}$Pb with respect to $^{10}$Be-n relative energy in the centre of mass of the projectile. Beam energy is 70 A.MeV. Full line: full-range bound state wave function. Dashed line: asymptotics.}
\label{as}
\end{figure}

 This formalism allows us therefore to directly compare calculations involving the full-range bound state wave function, calculated as shown in Section \ref{bou}, to those obtained with the asymptotic form of such wave function, as done previously \cite{mar02}. Fig. \ref{as} shows calculations of $^{10}$Be-n relative energy spectrum after Coulomb breakup of $^{11}$Be against a $^{208}$Pb target at an incident energy of 70 A.MeV \cite{fuku04}. The asymptotic approximation is indeed very close to the full-range wave function, and fully within experimental error bars, as will be shown below. Besides, the asymptotics yields a slightly larger value of the differential cross-section  in particular for the higher values of relative energy, as expected. The difference between the two approaches can be easily understood from a simple Hulth\`en-like argument. It must be pointed out, however, that this check can only be carried out for $s$-waves, because, in this formalism, the probability amplitude of eq. (\ref{nuq}) diverges upon integration over $\vec{k}$, when the asymptotic expressions for the wave functions are used and $\ell  > 0$, due to the shape of the Hankel functions involved. In Refs.\cite{mar02,mar03} it was discussed that up to $\ell =2$ this problem can be solved by calculating first the Fourier transform of the wave function in the $k_z$ direction and then getting the cross sections from the integral of the momentum distribution which is not divergent. On the other hand for any arbitrary  $\ell$, total cross sections are well calculated using a numerical wave function while the check of peripherality  can be done as discussed in the forthcoming Sec. 5.
\begin{figure}
\center
\includegraphics[scale=0.4,width=8cm]{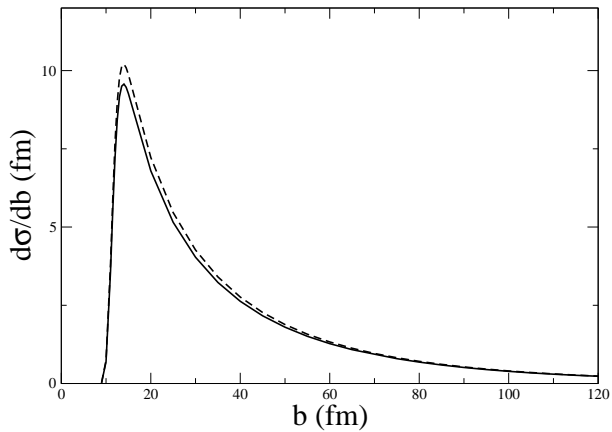}
\caption{Calculated impact parameter profile in Coulomb breakup of $^{11}$Be on $^{208}$Pb at 70 A.MeV. Full line: full-range bound state wave function. Dashed line: asymptotics.}
\label{as2}
\end{figure}

 From the point of view of the impact parameter dependence, a similar comparison can be carried out. Fig. \ref{as2} displays the impact parameter profile of the same reaction. The $^{10}$Be-n relative energy has now been integrated over, whereas the impact parameter is now along the x-axis. No cross-section is produced at low impact parameters due to the presence of the core survival probability $|S_{ct}(b)|^2$. The maximum is reached at a distance near the sum of core and target radii, and then the cross-section decays with a long tail. The difference between the full wave function and its asymptotic approximation decreases for larger values of impact parameter, as these imply the neutron being far away from the $^{10}$Be core, hence involving bigger values of $r$, for which the asymptotics are better.

\begin{figure}
\center
\includegraphics[scale=0.5,width=10cm]{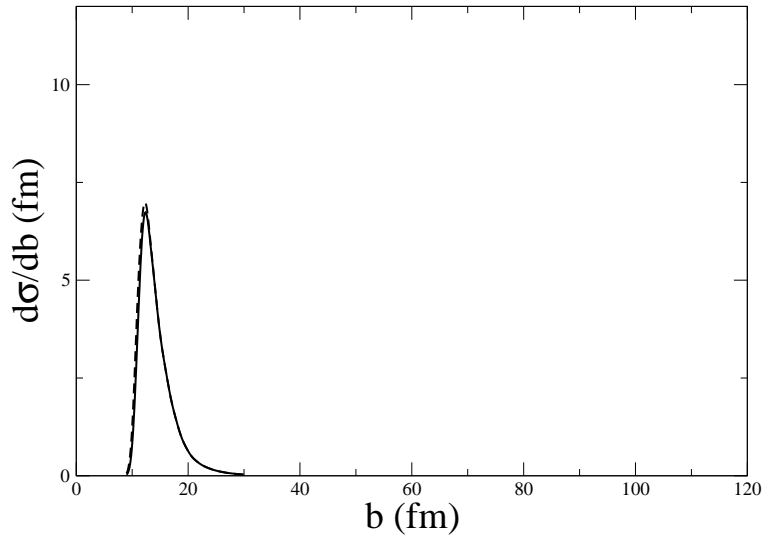}
\caption{Calculated impact parameter profile in nuclear elastic breakup of $^{11}$Be on $^{208}$Pb at 70 A.MeV. Full line: full-range bound state wave function. Dashed line: asymptotics.}
\label{as3}
\end{figure}

 When nuclear breakup of the same system is looked upon, equivalent features come out. Fig. \ref{as3} shows a similar impact parameter profile for this $^{11}$Be reaction, but now the breakup is caused only by the nuclear interaction. Again, or perhaps more clearly in this case, asymptotic approximation and full wave function disagree the most for the lower half of the relevant impact parameter range, yielding almost identical results for the upper one. As expected, the decay of the profile with impact parameter is much faster now than in the Coulomb breakup case, what obviously accounts for the long-range character of the latter. In any case, these calculations show that no important effect is introduced by the approximation of the wave function by its tail.

\subsection{Validity of the dipole approximation}
\begin{figure}
\center
\includegraphics[scale=0.5,width=10cm]{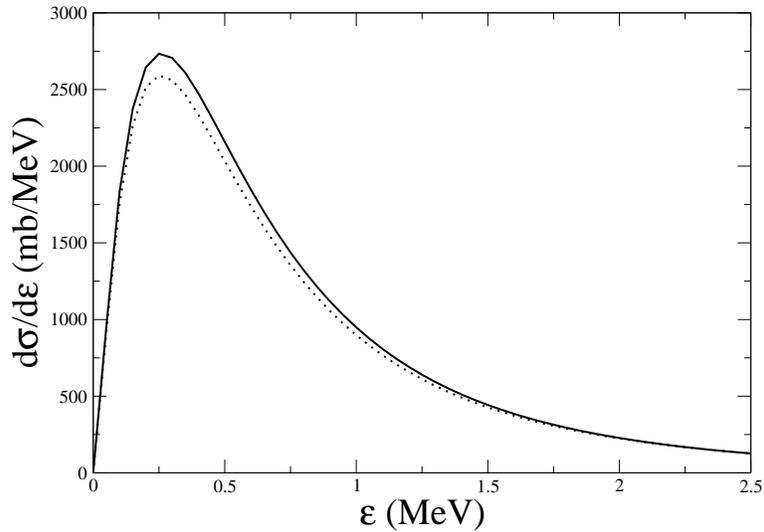}
\caption{Calculations of differential cross-section for Coulomb breakup of $^{11}$Be on $^{208}$Pb with respect to $^{10}$Be-n relative energy in the centre of mass of the projectile. Full line: dipole approximation. Dotted line: full multipole expansion.}
\label{mul}
\end{figure}

  As discussed in section \ref{teo}, the phase given by eq. (\ref{fiqui}) contains the Coulomb interaction to all orders in the multipole expansion. Thus the importance of high order multipole terms can be checked by comparing the results obtained with such expression with those provided by the first order term in its multipole expansion, i.e. the dipole phase eq. (\ref{fuqui}). In Fig. \ref{mul} such comparison is performed. As expected from earlier works, high order effects are of little importance in this case. In addition, they interfere destructively with the dominant dipole term, giving rise to a decrease in the single particle cross section, as has been also shown previously \cite{mar02,mar03}.

\subsection{All orders and nuclear effects}
\begin{figure}
\center
\includegraphics[scale=0.5,width=9cm]{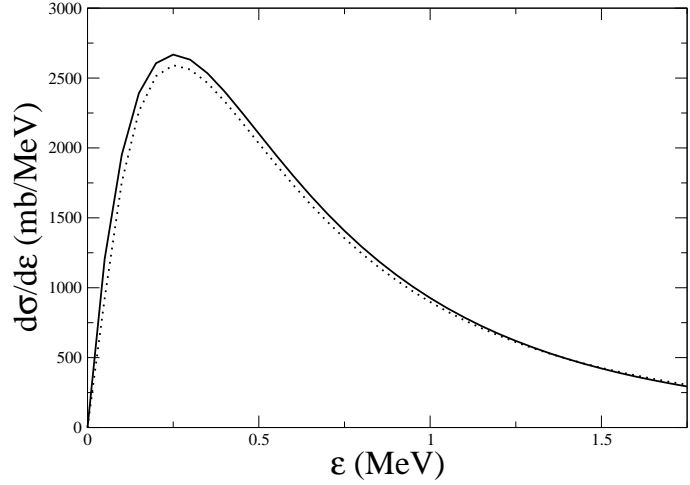}
\caption{\small Differential cross section as a function of core-target relative energy for Coulomb breakup of $^{11}$Be on $^{208}$Pb at 70 A.MeV. Full line: all orders. Dotted line: first order time-dependent perturbation theory. }
\label{addd}
\end{figure}
\begin{figure}
\center
\includegraphics[scale=0.5,width=9cm]{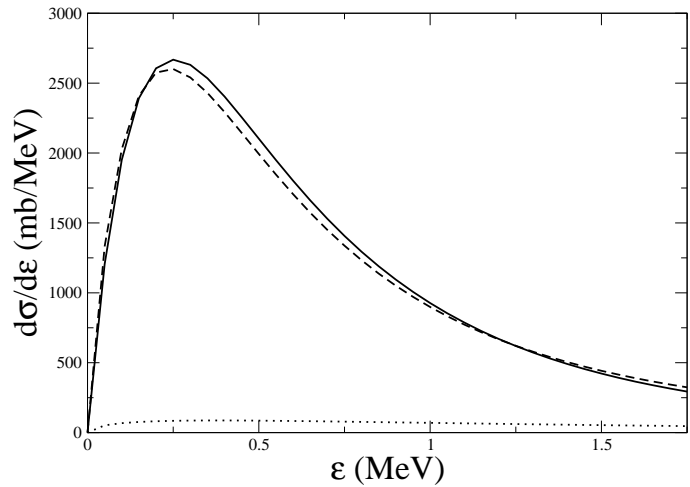}
\caption{\small Differential cross section as a function of core-target relative energy for Coulomb breakup of $^{11}$Be on $^{208}$Pb at 70 A.MeV. Full line: all-multipole-all-order    calculation for Coulomb breakup. Dashed line: including the nuclear interaction in the sudden approximation. Dotted line: nuclear breakup \cite{mar02}.}
\label{fffuuu}
\end{figure}

 As an extension of the work performed in \cite{mar03}, once the inclusion of all the terms in the multipole expansion of the potential has been achieved within this formalism, the effect of higher order terms in the strength of the potential is assessed next. In \cite{mar03} it was shown that a possible way to take into account this higher order strength terms is to use the  sudden approximation in which the (impact parameter divergent) first term is subtracted, and then a first order term, calculated in time-dependent perturbation theory, is added. This technique is now revisited with our full-multipole version expressions for Coulomb phase and its sudden approximation, eqs. (\ref{fiqui}) and (\ref{suddi}) respectively. Moreover, the full-range bound state wave function is used instead of its tail, although this effect has been above proved to be small. Fig. \ref{addd} shows the result. Again, a first order approach seems to suffice to account for the vast majority of the physics involved.

 Finally, notwithstanding the fact the elastic breakup of $^{11}$Be on Pb is well-known to be Coulomb dominated, the nuclear breakup are included in this all-order multipole-and-strength calculation, and the result showed in Fig. \ref{fffuuu}. Unsurprisingly, the inclusion of nuclear breakup makes very little alteration to the only-Coulomb result.

\subsection{Comparison to data}
\begin{figure}
\center
\includegraphics[scale=0.5,width=8cm]{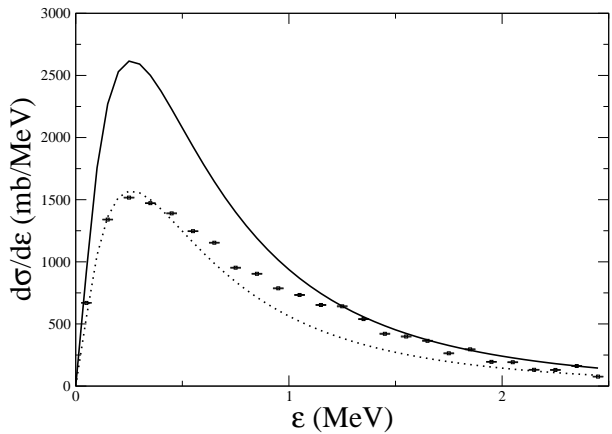}
\caption{Our calculation for the relative energy spectrum following $^{11}$Be elastic breakup compared to full angular coverage measurements. Full line: theoretical calculation with spectroscopic factor 1. Dotted line: theoretical calculation scaled with 0.6 to match the peak of the experimental data. Squares: data from \cite{fuku04}.}
\label{com}
\end{figure}
\begin{figure}
\center
\includegraphics[scale=0.5,width=8cm]{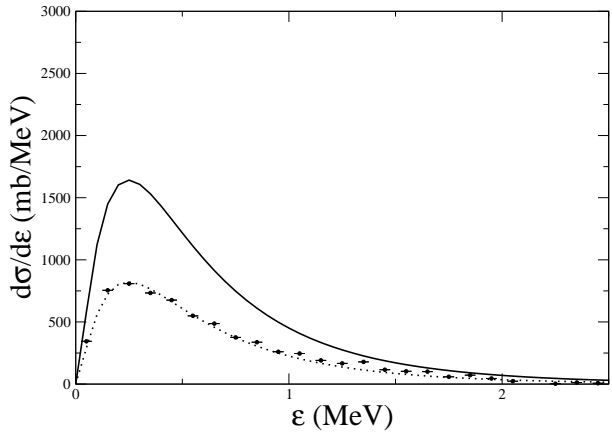}
\caption{Forward-angle relative energy spectrum for elastic breakup of $^{11}$Be. Full line: theoretical calculation with spectroscopic factor 1. Dotted line: theoretical calculation scaled with 0.5 to match the peak of the experimental data. Squares: data from \cite{fuku04}.}
\label{forw}
\end{figure}

 Let us now turn our attention to the matter of how these calculations perform when compared to experimental data. Figs. \ref{com} and \ref{forw} explore this issue. If the whole angular range measurements are taken (Fig. \ref{com}), the spectroscopic factor suggested to match the peak of the experimental data would be 0.6. However, the shape of the distribution is poorly described, what shed some suspicions about the suitability of our method to calculate this particular observable. On the other hand, Fig. \ref{forw} shows the same comparison for forward-angle measurements. In order to match this requirement, the usual semiclassical relation between scattering angle and impact parameter has been used \cite{fuku04}, in such a way that a minimum impact parameter is given for a maximum scattering angle. We have taken this maximum impact parameter value as 20 fm. When this is carried out, the peak of the data is reached by the inclusion of a factor 0.5. Besides, the shape of our calculation agrees now much better with that of the entire experimental distribution. This is not surprising as both in the eikonal approximation and in perturbation theory we use a straight line trajectory which works at best in processes happening at high impact parameter, and where high-order, multi-step processes are of little relevance. A similar result has been obtained by the numerical solution of the neutron Schr\"odinger equation method of Baye and collaborators \cite{capel04}. 
 
 The analysis in this section has been performed for a typical halo nucleus at an energy at which experiments have been performed
 at various laboratories. Another interesting projectile to study \cite{bon99} would be $^{19}$C whose initial state wave function contains also d-components and whose experimental neutron separation energy is not perfectly determined. Also, for this nucleus the numerical solution of the Sch\"odinger equation is probably quite complicated. In the next section we will indeed study the sensitivity of our results to  separation energies and to the initial angular momentum for heavy projectiles.
As far as the incident energy is concerned our conclusions are certainly valid at higher energies, while  incident energies close to the Coulomb barrier cannot be treated by the eikonal approximation.

\section{Tightly bound systems}

 For non-halo systems, this formalism can be used to investigate the general trend of nuclear and Coulomb  breakup cross-section as a function of the separation energy. The projectile under study is now $^{34}$Si whose valence d-orbital neutron has a separation energy 
  S$_n$=7.54 MeV. Experimental data already exist \cite{enders01}. From now on we shall include also stripping in our calculations since this component is dominant for normally bound nucleons. Before showing the results for total breakup cross sections we wish to investigate the question of the peripherality of the reaction.

   As mentioned early on, nuclear breakup is calculated in the sudden approximation, thus our formulae for elastic breakup and stripping cross sections, after averaging over angular momentum projections, are
\begin{eqnarray} \label{ele}
\sigma_{el}= \int d\vec{b} |S_{ct}(b)|^2 \int d\vec{r} |\phi_0(r)|^2 |1-S_{nt}(|\vec{b}+\vec{r}|)|^2
\end{eqnarray}
and
\begin{eqnarray} \label{este}
\sigma_{st}= \int d\vec{b} |S_{ct}(b)|^2 \int d\vec{r} |\phi_0(r)|^2 \left(1-|S_{nt}(|\vec{b}+\vec{r}|)|^2\right).
\end{eqnarray}
 
\begin{figure}
\center
\includegraphics[scale=0.35,width=8cm]{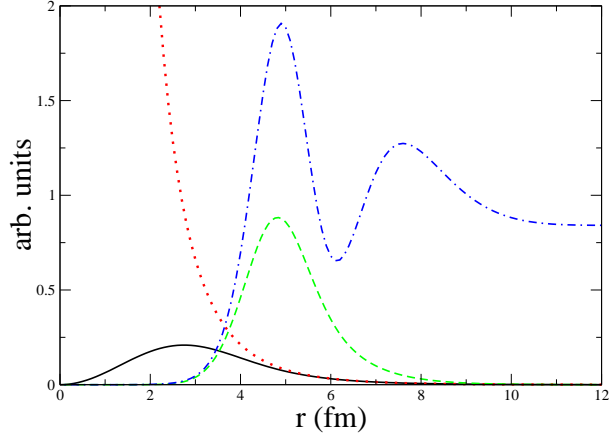}
\caption{Different factors entering the integration over $r$ in the calculation of the elastic breakup cross-section, eq. (\ref{ele}) for a core-target impact parameter of $b=12$ fm. Full line: radial wave function; dotted line: asymptotic form; dotted-dashed line: the factor $|1-S_{nt}(b-r)|^2$; dashed line: integrand of $r$-integration in eq. (\ref{ele}) magnified by a factor 50 to make it visible.}
\label{e12}
\end{figure}

\begin{figure}
\center
\includegraphics[scale=0.35,width=8cm]{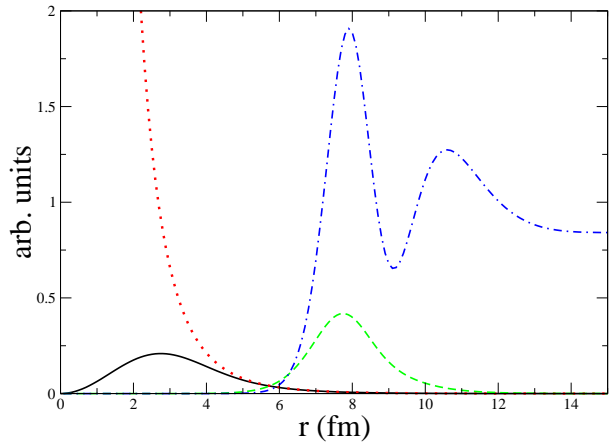}
\caption{Same as Fig. \ref{e12} for an impact parameter of 15 fm.}
\label{e15}
\end{figure}

\begin{figure}
\center
\includegraphics[scale=0.35,width=8cm]{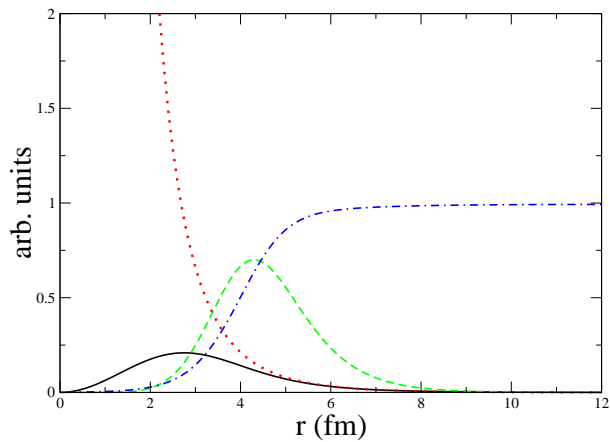}
\caption{Same as Fig. \ref{e12} but for stripping with a core target impact parameter of $b=12$ fm. Dotted-dashed line is now $1-|S_{nt}(b-r)|^2$. The magnifying  factor is now of 2000.}
\label{s12}
\end{figure}

\begin{figure}
\center
\includegraphics[scale=0.35,width=8cm]{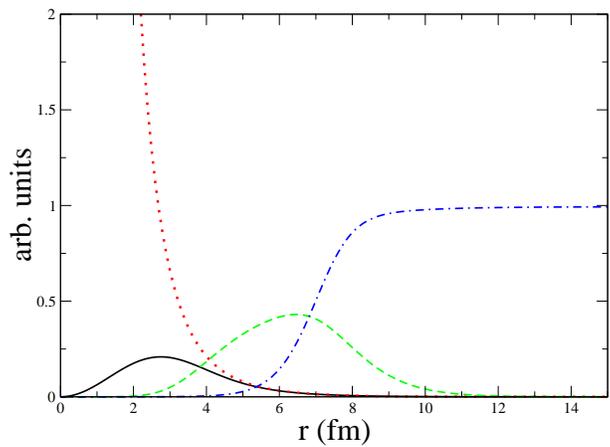}
\caption{Same as Fig. \ref{s12} for an impact parameter of 15 fm.}
\label{s15}
\end{figure}

 Here $S_{nt}(|\vec{b}+\vec{r}|)=e^{i\chi_{nt}(|\vec{b}+\vec{r}|)}$ defined by eq. (\ref{nuq}). In order to address peripherality, we study how different elements of the integral over $\vec{r}$ in above equations behave for a fixed value of impact parameter. We show them in Figs.\ref{e12}, \ref{e15}, \ref{s12} and \ref{s15}, for core-target impact parameters of $b=12$ fm and $b=15$ fm, two values close to that where cross section is maximum. The terms $|1-S|^2$ and $1-|S|^2$ have been projected onto the direction of $\vec{r}$ for simplicity. The asymptotic part of the wave function is also shown. This matches perfectly the wave function from 5 fm onwards. Similar figures for a much lighter system ($^{12}$Be+$^{9}$Be) where shown and discussed by Ref.\cite{bon01}. It is clear from these figures that although the integral in eqs. (\ref{ele}) and (\ref{este}) extends up to r=0, the integrand is concentrated at the surface. Furthermore smaller core-target separations are excluded by the experimental detection condition of a surviving core. At the same time the target, four times heavier than the projectile, will suffer a negligible excitation and no recoil.
 

\begin{figure}
\center
\includegraphics[scale=0.5,width=8cm]{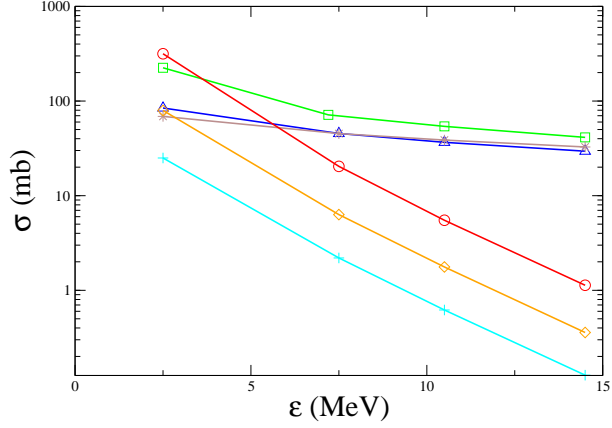}
\caption{Integrated  breakup cross-section for a hypothetical Si beam against Pb at 70 A.MeV as a function of the neutron separation energy. Different initial parameters: circles (squares) are for Coulomb (nuclear) breakup from an initial s-wave; diamonds (triangles) for Coulomb (nuclear) breakup from a d-wave; pluses (stars) for Coulomb (nuclear) breakup from an initial f-wave. Nuclear breakup is the sum of diffractive and stripping contributions}
\label{ens}
\end{figure}

Next,  Fig.\ref{ens} shows the result for the breakup cross sections. In the $^{34}$Si system, bound states with 2.5, 7.5, 10.5 and 14.5 MeV of separation energy have been artificially generated as s, d and f-waves, and subsequent  breakup cross-sections have been calculated using a Pb target at 70 A.MeV. As Fig. \ref{ens} displays, for the three angular momenta chosen the nuclear breakup seems to reach a limit value of separation energy, where cross-sections depend just on nuclear sizes. On the other hand, Coulomb breakup continues decreasing as separation energy increases, to a point in which, even for a heavy target, becomes negligible. The curve suggests a dependence ${\sigma}_{coul} \propto \exp{(- \epsilon_i/2)}$.

\begin{figure}
\center
\includegraphics[scale=0.5,width=9cm]{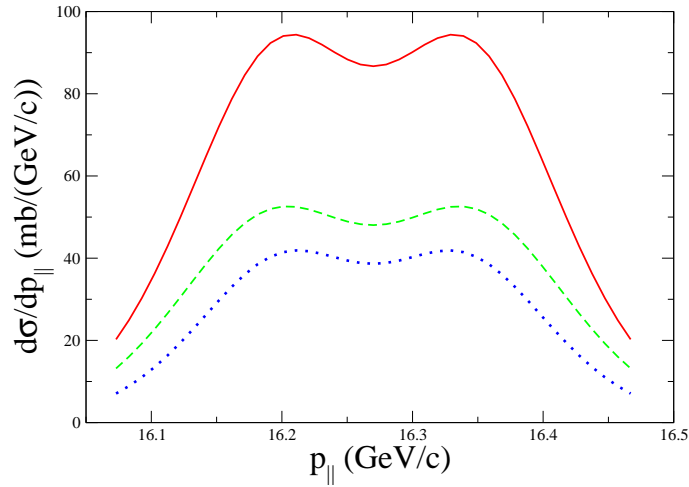}
\caption{Model prediction for the core parallel momentum distribution of $^{45}$Ar cores from an initial $\ell$=3 state. The stripping (elastic breakup) part is the dashed (dotted) line. Total is the full line.}
\label {sisitc}
\end{figure}

\begin{figure}
\center
\includegraphics[scale=0.5,width=9cm]{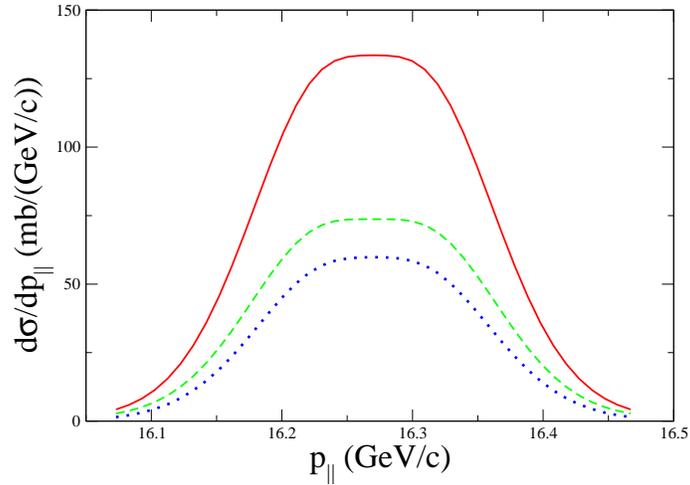}
\caption{The same core parallel momentum distribution as in Fig. \ref{sisitc} with an $\ell$=2 bound state. The stripping (elastic breakup) part is the dashed (dotted) line. Total is the  full line.}
\label {sisieik}
\end{figure}

\begin{figure}
\center
\includegraphics[scale=0.5,width=9cm]{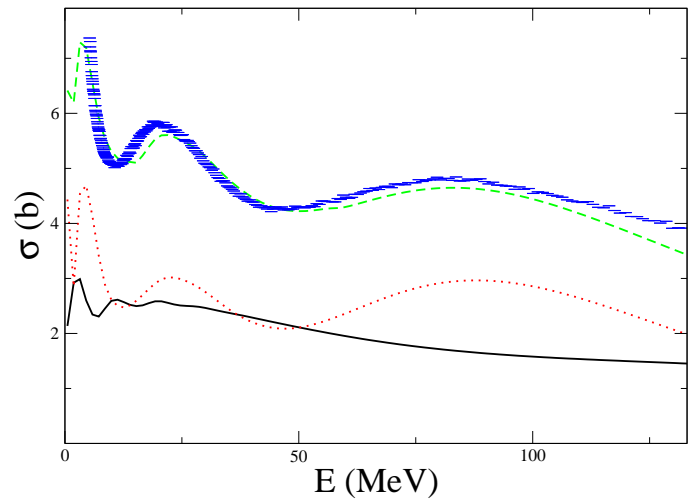}
\caption{Experimental total  n-$^{208}$Pb cross sections \cite{fin} and the calculated JLM \cite{jlm} elastic (dotted line), inelastic (full line) and total (dashed line) cross sections}
\label {npbpot}
\end{figure}

Finally on Figs.\ref{sisitc} and \ref{sisieik}   we show the core parallel momentum distributions after one neutron breakup (with S$_n$ = 8.07MeV, $\ell$=3 and  $\ell$=2 respectively) of $^{46}$Ar on Pb at 70 A.MeV. The eikonal total cross sections (full line)  calculations show the typical shapes characteristic of nuclear breakup
in the same way as previously seen for much lighter and less bound exotic projectiles on light targets. The corresponding cross sections are $\sigma_{st}$= 15.7 mb and
$\sigma_{el}$=12.0 mb for the initial d-state,
$\sigma_{st}$=16.0 mb and
$\sigma_{el}$=11.9 mb for the initial f-state. The curves include elastic breakup and absorption which are also separately shown in the same figures by the dotted and dashed lines respectively. This last example shows that the deviations from the eikonal model seen in the experimental data of Ref.\cite{gade05} could be overcome by using a heavy target, for which the peripherality and no-target-recoil hypothesis would hold.  For completeness we show also in Fig. \ref{npbpot} the experimental free n-$^{208}$Pb cross sections \cite{fin} and the calculated elastic, inelastic and total cross sections for the same system. We used the effective nucleon-nucleon interaction of Jeukenne, Lejeune and Mahaux (JLM) \cite{jlm1,jlm}, folded with the Pb density \cite{kara02} to obtain the n-Pb optical potential.


\section{Conclusions}
In this paper we have presented a - mostly analytical - method to take into account the effect of the full Coulomb potential to all orders in the interaction in the semiclassical scattering of two nuclei. The corresponding core-recoil effect for an exotic nucleus projectile described as a core plus neutron system, gives rise to the so called Coulomb breakup. We have then calculated Coulomb and nuclear breakup to all orders and checked the "peripherality" of the reaction for light and heavy targets and for halo and normally bound neutron rich nuclei. The figures show that the effects of the asymptotic approximation, higher multipoles and higher order in perturbation theory are accurate to within a few percent.

Finally we have shown that, for the medium mass to heavy exotic nuclei that have started to be studied recently and will be further produced in the future, the Coulomb breakup effect on a heavy target becomes negligible with respect to the nuclear breakup part. This happens because of the "normal" separation energies ( $\gtrsim $10MeV) and the large angular momentum of the valence orbitals $ \ell \geqslant$ 2. Therefore we have suggested to study these kind of nuclei by nuclear breakup reactions on heavy targets since all the conditions to apply the core-spectator model and the no recoil approximation for the target are satisfied.

\end{document}